\documentclass{aastex}
\usepackage{emulateapj5}
\usepackage{apjfonts}

\newcommand{\msun}{\mbox{M$_{\sun}$}}
\newcommand{\rsun}{\mbox{R$_{\sun}$}}
\def\etal{et al.\ }

\newcommand{\swin}{Centre for Astrophysics and Supercomputing, Swinburne
University of Technology, P.O.~Box 218 Hawthorn, VIC 3122, Australia}

\begin{document}

\title{Discovery of Two Relativistic Neutron Star--White Dwarf Binaries.}
\author{R. T. Edwards and M. Bailes}
\affil{\swin}

\begin{abstract}
We have discovered two recycled pulsars in relativistic orbits as part
of the first high-frequency survey of intermediate Galactic latitudes.
PSR J1157--5112 is a 44 ms pulsar and the first recycled pulsar with
an ultra-massive ($M > $~1.14~\msun) white dwarf companion.
Millisecond pulsar J1757--5322 is a relativistic circular-orbit system
which will coalesce due to the emission of gravitational radiation in
less than 9.5 Gyr. Of the $\sim$40 known circular orbit pulsars,
J1757--5322 and J1157--5112 have the highest projected orbital
velocities. There are now three local neutron-star/white-dwarf
binaries that will coalesce in less than a Hubble time, implying a
large coalescence rate for these objects in the local
Universe. Systems such as J1141--6545 \citep{klm+00a} are potential
gamma-ray burst progenitors and dominate the coalescence rate, whilst
lighter systems make excellent progenitors of millisecond pulsars with
planetary or ultra-low mass companions.

\end{abstract}


\section{Introduction}
The first surveys of the Galactic plane at high frequencies for radio
pulsars \citep{cl86,jlm+92} uncovered a much younger and more distant
population than the traditional 400 MHz surveys.  High frequency
surveys are much less affected by the deleterious effects of
scatter-broadening and dispersion smearing which make short period
pulsars impossible to detect at large dispersion measures.  
Unfortunately, the sampling rates employed in these early
surveys limited their sensitivity to millisecond 
pulsars (MSPs) \citep{jb91},
and the spatial coverage was limited by the small beam of a radio
telescope at high frequencies. No millisecond pulsars were
discovered in these early high-frequency surveys.
It was widely believed \citep{ffb91}
that MSPs were steep-spectrum objects, and that there
was little point in searching for them at high frequencies.  As the
population of MSPs became larger however, a series of spectral
studies \citep{lylg95,tbms98} provided evidence that 
spectra of millisecond
pulsars were not too dissimilar from that of normal pulsars. In a
simulation of the Galactic millisecond pulsar
population, \citet{tbms98} predicted that high frequency
surveys undertaken with the Parkes multibeam system \citep{swb+96}
should detect a large number of millisecond pulsars. Motivated
by this study we have conducted the first large-scale survey for
millisecond pulsars at intermediate Galactic latitudes.

  Our survey covered the region enclosed by 
$5\degr < |b| < 15\degr$ and $-100\degr < l < 50\degr$. 
Pointed observations of 265 seconds were made
with the 64-m Parkes radiotelescope using the 21-cm multibeam
receiver. 
The receiver consists of thirteen dual linear polarization
feeds arranged in a sparse focal-plane
pattern which allows full sky coverage with beams overlapping at the
half-power points. This increases the rate of sky coverage by a 
factor of thirteen over a single-beam receiver and enabled us to
conduct a large-scale survey in only 14 days of integration time.
The system is sensitive to frequencies from 1230
MHz to 1530 MHz, with an average total system noise temperature of 24
K. The backend system was built at Jodrell Bank for the ongoing deep
Galactic plane multibeam pulsar survey \citep{lcm+00}.  
It includes
twenty-six filterbanks, each with ninety-six channels and total
bandwidth of 288 MHz, centred at a sky frequency of 1374 MHz.
Filterbank outputs were summed in polarisation pairs and one-bit
digitised with an integration time of 125 $\mu$s. The large bandwidth,
narrow channels, fast sampling, low system temperature and multiple
beams facilitiated the discovery of a large number of pulsars,
including millisecond pulsars, with greater time efficiency than any
previous large survey.
Data were recorded on magnetic tape for offline processing on the
Swinburne supercluster, a network of 64 Compaq Alpha workstations, using
standard techniques \citep{mld+96}.

 The survey is now complete. Full details of the survey will be
provided in a subsequent paper (Edwards et al., in prep.). In all, we have
discovered 58 new pulsars, of which one has a period of 44 ms and
seven others have periods less than 20 milliseconds. In this {\it
Letter} we will discuss the two recycled pulsars for which we have
phase-connected solutions, PSRs J1157--5112 and J1757--5322.

\section{PSR J1157--5112 --- Ultramassive Companion}

PSR J1157--5112 is a 44~ms pulsar in a 3.5 d circular orbit,
discovered in an observation made on 1999 Jan
19th. Its parameters, derived from
pulse time-of-arrival analysis, are listed in Table 1. 
The pulsar's
relatively short pulse period, small spin-down rate
and circular orbit 
indicate recycling in a binary from an
evolved companion \citep{vdh84}. If the companion formed
a neutron star after recycling the pulsar, we would expect
an eccentric orbit due to the sudden mass loss associated
with the neutron star's production. The circularity of the
orbit implies that the companion is a white dwarf.
Since observed neutron star masses are consistent with a Gaussian
distribution with a standard deviation of just 0.04~\msun\ around the
mean of 1.35~\msun\  \citep{tc99},
a conservative lower limit on the mass of the companion white dwarf
can be obtained by assuming a 1.27~\msun\ pulsar and
an edge-on orbit. 
This means that $M_{\mathrm{WD}} >$~1.14~\msun, indicating an
``ultra-massive'' ONeMg degenerate, and could easily be
near the Chandrasekhar limit if the orbit is near the median
inclination angle of binaries.
By similar reasoning, if the white dwarf mass is
less than 1.35~\msun, the pulsar mass is at most 1.75~\msun.

A number of ultra-massive white dwarf systems are now known. The
largest sample are those that have been discovered in recent X-ray and
extreme ultraviolet observations \citep{bsl92,vtgd97},
the masses of which are estimated from model-dependent spectroscopic
measurements.  The companion to PSR J1157--5112 is the first
ultra-massive white dwarf to have a firm lower mass limit which
relies only on the narrow band of observed neutron star masses and
Newtonian physics.  Steller mergers have been suggested for the
formation of ultra-massive white dwarfs, however it is clear from the
maintenance of a binary association that this is not the case with the
companion of PSR J1157--5112, which must have resulted from largely
standard stellar evolution.

PSR J1157--5112 is the latest example of a pulsar system that
possesses a very massive white dwarf companion.  Recent optical
observations of the eccentric binary pulsar B2303+46 have 
indicated that a faint object
consistent with a white dwarf is coincident with the position of the
pulsar \citep{vk99}.  If this proves to be a genuine association, the
periastron advance and mass function limit the companion mass to
greater than 1.2 \msun.  The newly discovered pulsar binary
J1141--6545 \citep{klm+00a}, with a minimum companion mass of 0.95~\msun\ 
is probably also a white dwarf \citep{klm+00a,ts00}.  The
orbits of these systems are highly eccentric and thus it is
believed that the neutron stars were born {\it after} their white
dwarf companions. Since the pulsar in the PSR J1157--5112 binary is recycled,
it is the first example of an ultra-massive white dwarf 
that was formed after the neutron star.

The high mass ratio between any possible progenitor to the white dwarf
and the pulsar must have resulted in a phase of unstable mass
transfer.  The system probably passed through a common envelope phase
when the white dwarf progenitor was on the asymptotic giant branch
(AGB), as suggested by \citet{vdh94} for the PSR J2145--0750 system.
Recent models \citep{rgi99} indicate that main sequence stars as massive
as 11 \msun\ can avoid core collapse on the AGB to produce
ultra-massive white dwarfs ($M_{\mathrm {WD}} = 1.37$ \msun\ in the
above case). Given that envelope loss in the common envelope phase
reduces the resultant remnant mass from a donor star of given
mass \citep{it93}, we expect that the progenitor to the white dwarf
companion of PSR J1157--5112 weighed in excess of 10 \msun.  This
would mean that the companion's radius \citep{rgi96,gri97a,irg97,rgi99}
was $<800$ \rsun\ at the time of Roche lobe overflow. The orbital
energy lost during spiral-in to the current orbit was insufficient to
expel the envelope of the companion, without an additional energy
source such as an accreting neutron star.  Future proper motion
measurements will help constrain models for its evolution.  For an 11
\msun\ companion with a pre-CE Roche lobe radius $>$ 500 \rsun\
(requiring that at most 7/8 of the envelope ejection energy comes from
non-orbital sources), the kick velocity of the pulsar birth event
must have been $<$~120~km~s$^{-1}$ for the system to have
remained bound. On the other hand, if the system was much closer at
the time of explosion much greater kicks could be accommodated and we
might expect the system to have a very large runaway
velocity \citep{bai89}.

Along with J1757--5322 (below), J1157--5112 brings the total number of
intermediate mass binary pulsars (those with white dwarf companions
heavier than 0.6~\msun) \citep{acw99,ts99a,cnst96} to seven. The
companion of PSR J1157--5112 is by far the most massive of the class.
That the neutron star survived what seems to have been a common envelope
phase with a massive AGB star strongly supports the case 
against black hole
formation via hypercritical accretion \citep{al00}.

\section{PSR J1757--5322 -- Coalescing Binary}

PSR J1757--5322 is an 8.9 ms pulsar which was
 discovered in an observation made on 1999 May 8th,
at an epoch where the orbital acceleration experienced by the pulsar
($\sim$7.5 km s$^{-2}$ in the line-of-sight) resulted in noticable
period evolution in the 265-s integration.
Subsequent observations revealed the pulsar to be in a nearly circular
11-h orbit with a companion of at least 0.55 \msun. A phase coherent
timing solution has been obtained, the parameters of which are listed
in Table 1. 
From similar reasoning to that invoked above
for J1157--5112, we expect that the companion is a white dwarf. Its
parameters are compatible with the common envelope model discussed
above for the intermediate mass binary pulsars.

The J1757--5322 system is highly relativistic, having the highest
orbital velocity of all circular orbit binaries (followed by
J1157--5112). The near-circularity of the orbit reduces the
measurability of relativistic effects,
however some effects could be measurable in the coming decades with
long timing baselines.  Specifically,
assuming an inclination angle $i=60\degr$ (the median value for
randomly oriented orbits), General Relativity predicts that the
periastron of J1757--5322 is advancing at a rate of
$\sim$~1\fdg2~yr$^{-1}$, that its orbital period is decreasing by
1.8~$\mu$s~yr$^{-1}$, that the pulsar's spin axis is precessing an
entire turn every 1700 yr and that the system will coalesce in at most
9.5~Gyr due to the emission of gravitational waves.  The Shapiro
delays due to space-time curvature around the companion (for $i=60\degr$)
are 29~$\mu$s
and 13~$\mu$s peak-to-peak respectively for PSR J1157--5112 and PSR
J1757--5322, and may be considerably longer if the systems
are nearly edge-on to the line of sight.

 After about 7.6 Gyr the orbital period of J1757--5322 will have
decreased to one-third its current value, resulting in four times the
orbital acceleration. Such a system, if observed in our survey would
have experienced significant signal-to-noise ratio loss, limiting the
chance of detection. However, such a system will only live for
$\sim$400 Myr, as compared to $\sim$8 Gyr for the preceding less
accelerated post-CE evolution phase in which J1757--5322 was
detected. Such highly accelerated systems are thus likely to be an
order of magnitude lower in detectable population.

The companion of J1757--5322 will fill its shrinking Roche lobe when
it has an orbital period of $\sim$50~s. The range of potential
outcomes of this is diverse and interesting.  Models suggest that
after a period of mass transfer, the white dwarf will tidally disrupt
and form an accretion disk around the pulsar in just a few seconds if
the companion mass is greater than about 0.7 \msun  \citep{vb84}. The
rapid mass transfer will result in an enormous release of energy,
potentially in the form of a gamma-ray burst. However, it is
not clear whether or not the accretion luminosity would be
self-limiting due to the outward radiation pressure inhibiting mass
transfer  \citep{fwhd99}. If there is no remnant of the white dwarf we
might expect a solitary millisecond pulsar with a very rapid rotation
rate \citep{vb84} (possibly like PSR B1937+21) and bigger than average
mass.  Alternatively it may be that the mass of the pulsar is driven
over the critical limit for neutron stars and becomes a black hole. If
the disk surrounding the neutron star is not completely destroyed the
formation of a planetary system around a millisecond pulsar is not
unreasonable, similar to PSR B1257+12. On the other hand if just 5\%
of the mass of the white dwarf is retained intact, a short-period
eclipsing pulsar system such as PSR J2051--0827 may result.  The
measurement of the proper motion of the system is therefore vital in
addressing these issues. A large velocity would provide a good match
to the planet pulsar PSR B1257+12. A very small velocity is probably
less conclusive.

The observed rate of gamma-ray bursts in the local Universe
 \citep{phi91} is now broadly compatible with the expected coalescence
rate of pulsar-white dwarf binary systems.  There are now three such
systems known that will coalesce in the lifetime of the universe.  PSR
J0751+1807 will coalesce in $\sim$7 Gyr.  It is only 2 kpc away but
has a light 0.15~\msun\ companion that upon reaching its critical
Roche lobe will undergo stable mass transfer to its much heavier
pulsar companion, resulting in an ultra-compact X-ray
binary \citep{elc97}.  PSR J1141--6545, as mentioned above, has a white
dwarf companion of at least 0.95~\msun\ in an eccentric orbit that
will coalesce in only $\sim$1.3 Gyr, and is $\sim$3.2 kpc distant. It
is a prime candidate for rapid mass transfer after the onset of Roche
lobe overflow and a potential gamma-ray burst progenitor due to its
high mass.  The alarming thing about this pulsar is its youth. Its
characteristic age is just 1 Myr and it may not pulse for much more
than 10 Myr due to magnetic dipole breaking pushing it past the
rotation limit for the emission of radio pulses.  If we take the
standard pulsar beaming fraction of $\sim$5, and the short lifetime of
the binary in an observable state ($\sim$10 Myr) compared to the
coalescence time (1.3 Gyr), the extrapolated Galactic coalescence rate
for such objects is enormous! We could reasonably expect there to be
$5\times 1300/10=850$ dead white dwarf neutron star binaries within a
few kpc of the Sun that will coalesce in less than a Hubble time.  The
total Galactic population would then be 1-2 orders of magnitude
greater than this, and the coalescence rate near 10$^{-5}$ yr$^{-1}$
per Galaxy, similar to that expected of double neutron star binaries,
another candidate for gamma-ray bursts  \citep{phi91,npp92}.

The contribution of PSR J1757--5322 to the inferred deathrate of white
dwarf-neutron star binaries is much smaller.  With the relative
proximity ($\sim$2~kpc) but enormous characteristic age ($\sim$5~Gyr)
of PSR J1757--5322, we might expect $\sim$250 similar coalescing
systems in the Galaxy, with a coalescence rate of
5$\times10^{-8}$yr$^{-1}$ if we assume an evenly distributed disk
population 15~kpc in radius. Hence there should be $\sim$500 systems in
the Galaxy which have already coalesced, compatible with the observed
population of Galactic disk pulsars with planetary or evaporating
white dwarf remnant companions (three within 1.5~kpc).

  The white dwarf companions of PSR J1157--5112 and PSR J1757--5322
are possibly observable with 10m class telescopes or HST. PSR B0655+64
($M_{\rm WD} >$~0.64~\msun) has a timing age of 3.6 Gyr and a 22nd
R-magnitude companion \citep{kul86} at a distance of about 500 pc. The
pulsars described here are probably 2-4 times as distant and of comparable
timing ages so we might expect their companions to have R-magnitudes
of 24-26. Detections of the companions to PSR J1757--5322 and PSR
J1157--5112 will provide invaluable information about the cooling
times of massive white dwarfs as the pulsars give an independent
estimate of their age.

\acknowledgements
We thank Simon Johnston for helpful comments on the manuscript, and
the members of the Swinburne pulsar group (in particular W. van Straten)
for assistence with observations. We are indepted to the Galactic plane
multibeam survey team for the use of equipment built for that survey
and for timing observations made on our behalf. This work was supported
by an APA scholarship (R.E.) and ARC fellowships (M.B.) and grants.


\begin{deluxetable}{lrr}
\tablecaption{Astrometric, Spin, Binary and Derived Parameters}
\tablecolumns{3}
\tablewidth{0pt}
\tablehead{
    \colhead{} 
      & \colhead{J1157--5114}
      & \colhead{J1756--5322}
}
\startdata
Right ascenscion, $\alpha$ (J2000.0)\dotfill 
   & 11$^{\mathrm h}$57$^{\mathrm m}$08\fs166(1)
   & 17$^{\mathrm h}$57$^{\mathrm m}$15\fs1618(4)
\\
Declination, $\delta$ (J2000.0)\dotfill
& -51\degr12\arcmin56\farcs14(3)
& -53\degr22\arcmin26\farcs38(1)
\\
Pulse period, $P$ (ms)\dotfill
   & 43.58922706284(12)
   & 8.869961227275(4)
\\
$P$ epoch (MJD)\dotfill
    & 51400.000000
    & 51570.000000
\\
Period derivative, $\dot{P}$ (10$^{-20}$)\dotfill
   & 14.6(16)
   & 2.78(15)
\\
Dispersion Measure (pc cm$^{-3}$)\dotfill
   & 39.67(3)
   & 30.793(4)
\\
Orbital Period, $P_{orb}$ (d)\dotfill
   & 3.50738639(3)
   & 0.4533112382(7)
\\
Projected semi-major axis, $a \sin i$ (lt-s)\dotfill
   & 14.28634(3)
   & 2.086526(5)
\\
Epoch of Ascending Node, $T_{\rm asc}$ (MJD)\dotfill
   & 51216.4442642(14)
   & 51394.1080692(3)
\\
$e \cos \omega$\tablenotemark{a}\dotfill
   & -0.000322(4)
   & -1.0(40)$\times 10^{-6}$
\\
$e \sin \omega$\dotfill
   & 0.000240(4)
   & 4.3(45)$\times 10^{-6}$
\\
\cutinhead{Derived Parameters}
Longitude of periastron, $\omega$ (\degr)\dotfill
   & 306.7(6)
   & 347(58)
\\
Orbital eccentricity, $e$\dotfill
   & 4.02(4)$\times 10^{-4}$
   & 4.4(45)$\times 10^{-6}$
\\
Companion mass, $M_{\rm WD}$ (\msun)\dotfill
    & $>$1.14
    & $>$0.55
\\
Characteristic age, $\tau_c$ (Gyr)\dotfill
    & 4.70
    & 5.04
\\
Surface magnetic field, $B_{\rm surf}$ (10$^8$ Gauss)\dotfill
    & 25.55
    & 5.02
\\
\enddata

\tablenotetext{a}{We used the ELL1 binary timing model of TEMPO (Wex,
unpublished). To avoid the large covariance between the time of
periastron ($T_0$) and $\omega$ for $e \ll 1$, the time of ascending
node ($T_{\rm asc}$) and the Laplace parameters $e \cos \omega$ and $e
\sin \omega$ are used instead.}

\tablecomments{Values in parentheses apply to the final digit of the value
and represent twice the formal uncertainties produced by TEMPO after 
scaling TOA uncertainties to achieve a reduced $\chi^2$ of unity.}

\end{deluxetable}

\end{document}